\documentclass[sigplan, nonacm]{acmart}
\AtBeginDocument{%
  \providecommand\BibTeX{{%
    \normalfont B\kern-0.5em{\scshape i\kern-0.25em b}\kern-0.8em\TeX}}}

\copyrightyear{2022}
\acmYear{2022}
\setcopyright{rightsretained}
\acmConference[DAC '22]{Proceedings of the 59th ACM/IEEE Design Automation Conference (DAC)}{July 10--14, 2022}{San Francisco, CA, USA} \acmBooktitle{Proceedings of the 59th ACM/IEEE Design Automation Conference (DAC) (DAC '22), July 10--14, 2022, San Francisco, CA, USA} \acmDOI{10.1145/3489517.3530553} 
\acmISBN{978-1-4503-9142-9/22/07}

\usepackage{algorithmic}
\usepackage{textcomp}
\usepackage[linesnumbered,ruled,noend]{algorithm2e}

\SetCommentSty{mycommfont}

\usepackage{float}
\usepackage{bm}

\begin{document}

% \title{HD Hashing: Robust and Efficient Dynamic Hash Table Using Hyperdimensional Computing}
\title{Hyperdimensional~Hashing: A~Robust~and~Efficient~Dynamic~Hash~Table}


\author{Mike Heddes, Igor Nunes, Tony Givargis, Alexandru Nicolau and Alex Veidenbaum}
\affiliation{
  \institution{Department of Computer Science, University of California, Irvine}
  \city{Irvine}
  \state{California}
  \country{United States of America}
}
\email{{mheddes, igord, givargis, nicolau, alexv}@uci.edu}

\renewcommand{\shortauthors}{Heddes and Nunes, et al.}

\begin{abstract}
Most cloud services and distributed applications rely on hashing algorithms that allow dynamic scaling of a robust and efficient hash table. Examples include AWS, Google Cloud and BitTorrent. Consistent and rendezvous hashing are algorithms that minimize key remapping as the hash table resizes. While memory errors in large-scale cloud deployments are common, neither algorithm offers both efficiency and robustness. Hyperdimensional Computing is an emerging computational model that has inherent efficiency, robustness and is well suited for vector or hardware acceleration. We propose Hyperdimensional (HD) hashing and show that it has the efficiency to be deployed in large systems. Moreover, a realistic level of memory errors causes more than 20\% mismatches for consistent hashing while HD hashing remains unaffected.
\end{abstract}




\maketitle

\section{Introduction}
\label{sec:intro}
An important problem in many cloud services and distributed network applications is the process of mapping requests to available resources. Example systems include: load balancing in cloud data centers, web caching, peer-to-peer (P2P) services, and distributed databases. Difficulty arises in such highly dynamic systems because resources join and leave the cluster at any time, due for example to cloud elasticity~\cite{al2017elasticity}, server failures, or availability of peers in a P2P network. It is often desirable to distribute requests evenly among resources and to minimize the number of redistributed requests when a resource joins or leaves. A non-uniform mapping results in overloading of resources and critical failure points.

% In many network systems, storing items in a way that makes them easy to find later is one of the most important problems. A common strategy is to use a hash table, a data structure that allows searching in $\mathcal{O}(1)$ by creating a key-value map. However, in many real-world applications, such as cloud load balancing, web caching, and peer-to-peer (P2P) services, it is common for information to be stored in a dynamic, distributed system. In such cases, in addition to the need to distribute the data evenly (avoiding overloading and critical failure points), the servers that store the data can join or leave the system at any time \igor{exemplify with cloud elasticity/server failures}. 

The simplest hash table solves the mapping problem using modular hashing. Despite having a great lookup time complexity of $\mathcal{O}(1)$, a change in table size (number of available resources) requires virtually all requests to be redistributed due to the modulo operation (more details in Section~\ref{sec:background}). \textit{Consistent hashing}~\cite{karger1997consistent} and \textit{rendezvous hashing}~\cite{thaler1998using} are alternative hashing algorithms that minimize redistribution when the hash table is resized. They prevent resource overloading at the cost of increased lookup time---$\mathcal{O}(\log n)$ and $\mathcal{O}(n)$ respectively.

% In traditional hash tables, using modular hashing, a change in table size (number of servers) requires virtually all objects to be remapped \igor{ref?}, which application-wise usually means a massive disruption because of bandwidth-intensive data movement. \textit{Consistent hashing}~\cite{karger1997consistent} and \textit{Rendezvous hashing}~\cite{thaler1998using} are hashing techniques proposed as alternatives to traditional hash tables for this type of scenario as they allow for minimal disruption when the table is resized.

However, we show that when considered in a dynamic environment subject to errors and failures (i.e., noise), the performance of consistent hashing and rendezvous hashing in minimizing the number of redistributed requests degrades. Noise can be introduced in many aspects of a system. We focus on memory errors which can for instance be caused by soft errors in the form of single event upsets (SEU), multi-cell upsets (MCUs) or hard errors~\cite{hwang2012cosmic, sridharan2012study}. MCUs, or burst errors, occur during a single event and are becoming more common as the feature size decreases. For 22~nm technology MCUs are estimated to be 45\% of all SEUs~\cite{ibe2010impact}. Moreover, analysis of memory failures in Google's data centers revealed that each year a third of the machines experiences a memory error~\cite{schroeder2009dram}. More robust hashing alternatives make it possible for cloud providers to perform fewer memory swaps, reducing operation cost.

Hyperdimensional  Computing  (HDC) is an inherently robust emerging computational model developed by Kanerva~\cite{kanerva2009hyperdimensional} inspired by neuroscience. HDC tries to emulate brain-like computing by representing information using high-dimensional random vectors, called \textit{hypervectors}. This representation shares qualities from biological neural systems such as robustness and efficiency. Representation and  transformations of data in HDC are performed over hypervectors of fixed dimensions, allowing for massive parallelism. 

Fueled by the demonstrated properties of HDC and the aforementioned limitations of current hashing algorithms, we propose \textit{Hyperdimensional (HD) hashing}, a new HDC-based dynamic hashing algorithm. HD hashing scales similarly to consistent hashing while proving to be much more efficient than rendezvous hashing. HDC's highly parallelizable operations have been exploited in recent research, showing that special hardware can make HD hashing far superior in efficiency (more details in Sec.~\ref{sec:HDC}). Moreover, we show that our algorithm is significantly more robust against noise. With 512 servers and a 10-bit MCU, HD hashing is unaffected while rendezvous and consistent hashing mismatch 4\% and 12\% of requests, respectively. With MCUs becoming more common this poses a risk for critical failures.

Our second contribution is a novel HDC encoding for representing a circle in hyperdimensional space, we call these \textit{circular-hypervectors}. They are a core component of HD hashing as they provide the mechanism for mapping requests to servers.

% HD hashing employs periodic-hypervectors to homogeneously distribute requests across a variable population of servers and is at the same time robust and efficient. \igor{We show through our experiments...(how does it compare to the other two)}

% An important aspect of HDC is the process of mapping information onto hypervectors, called \textit{encoding}. Several encoding techniques have been proposed that show the usefulness of HDC in several applications. However, despite the great relevance of periodic/circular information \igor{such as ...}, none of the existing encoding techniques is capable of translating this feature in the hyperspace properly. With this motivation, the first contribution of this work is a methodology for generating \textit{periodic-hypervectors}, a set of circularly correlated basis-hypervectors. In addition to being the central component of our proposed hashing system which will be described below, in Section~\ref{sec:other_uses} we show through experiments periodic-hypervectors can be used in other HDC applications, such as machine learning, to improve accuracy over existing encoding methods \igor{mention how much improvement}.

\section{Background}
\label{sec:background}

\subsection{Consistent Hashing}
\label{sec:consistent_hashing}
Consistent hashing is a common way of distributing requests among a changing population of servers~\cite{mirrokni2018consistent,stoica2003chord} (often times, the problem and the technique are referred to as \textit{consistent hashing} indistinctly). The algorithm, which gave rise to Akamai~\cite{nygren2010akamai}, is used in many other real-world large scale applications such as Dynamo on Amazon Web Services~\cite{decandia2007dynamo} and Google Cloud Platform~\cite{eisenbud2016maglev}.

To describe consistent hashing, let $h(\cdot)$ denote a hash function that takes requests as inputs (in practice an IP address or unique identifier, for example) and $S=\{s_1 , \dots , s_n \}$ a set of servers. In modular hashing, a request $r$ is simply assigned to $s_i$ where $i=h(r)\bmod n$. Instead, consistent hashing maps both requests and servers uniformly to the unit interval $[0,1]$, which is interpreted as a circular interval. Thereafter, each request is assigned to the first server that succeeds it in the circle in clockwise order. This assignment is usually done in $\mathcal{O}(\log n)$ time using binary search.

\subsection{Rendezvous Hashing}
\label{sec:rendezvous_hashing}
The basic idea of rendezvous hashing~\cite{thaler1998using}, also known as highest random weight (HRW) hashing, is very simple. Given a hash function $h(\cdot)$ that takes as input a server and a request, each request $r$ is assigned to the server $s_i$ where:
\begin{align*}
    s_i =\underset{s\in S}{\arg\max} \;h(s,r)
\end{align*}
Each assignment is therefore done in $\mathcal{O}(n)$ time, since it is necessary to compute the hash of the request paired with each server in the system in order to compute the maximum value. In practice, Rendezvous hashing is used less often than consistent hashing, despite distributing the requests more uniformly, because of the increased time complexity.

\subsection{Hyperdimensional Computing}
\label{sec:HDC}

From a comparative study of computing in animal brains and computer logic circuits~\cite{kanerva2009hyperdimensional}, Hyperdimensional Computing (HDC) emerged as a robust and efficient alternative computation model. The central observation is that large circuits are fundamental to the brain's computation. HDC incorporates this notion by computing with 10,000-bit words (hypervectors), instead of 8-to-64-bit.

Such hyperspaces (short for hyperdimensional spaces) have properties that explain certain rich brain properties that are otherwise difficult to reproduce on computers. For example, hypervectors encode information holographically, meaning that each of the thousands of bits contains the same amount of information, ensuring inherent robustness~\cite{kanerva2009hyperdimensional, wu2018brain}. 

% HDC literature describes two types of hypervectors, random and level-hypervectors. Random hypervectors are sampled independently and uniformly, making them quasi-orthogonal to each other. This representation is used to describe categories because categories are often viewed as unrelated variables. Level-hypervectors are used to represent scalar values because it captures the relation of similar values having similar hypervectors.

In addition to representation, the other crucial part of a computer system is information manipulation, or arithmetic. The arithmetic in HDC is based on well-defined operations between hypervectors, such as addition (bundling), multiplication (binding) and permutation. Another important function is information comparison, which in HDC usually means measuring the similarity between hypervectors using the inverse Hamming distance or the cosine similarity. All those operations are typically dimension-independent, providing an opportunity for massive parallelism~\cite{li2016hyperdimensional,rahimi2017high}. 

Computational efficiency is one of the core motivations aimed at since the conception of HDC and it is envisioned for and expected to reach full potential in specialized hardware~\cite{kanerva2009hyperdimensional}. In addition to the just mentioned parallelizability, optimizations such as in-memory processing promise to further increase the computational efficiency of HDC~\cite{imani2017ultra}. Schmuck et al.~\cite{schmuck2019hardware} apply a series of hardware techniques to optimize HDC, such as on-the-fly \textit{rematerialization} of hypervectors and special memory architectures, to improve chip area and throughput at the same time.
Particularly important to substantiate the claims we make in this paper about efficiency (see Section~\ref{sec:HD_consistent_hashing}), they demonstrate an FPGA implementation that uses deep adder trees to perform inference in a single clock-cycle.
% \section{Distributed Hash Table Using Hyperdimensional Computing}
\section{Hyperdimensional Hashing}
\label{sec:HD_consistent_hashing}

\begin{figure}[ht]
 \centering
 \includegraphics[width=.75\columnwidth]{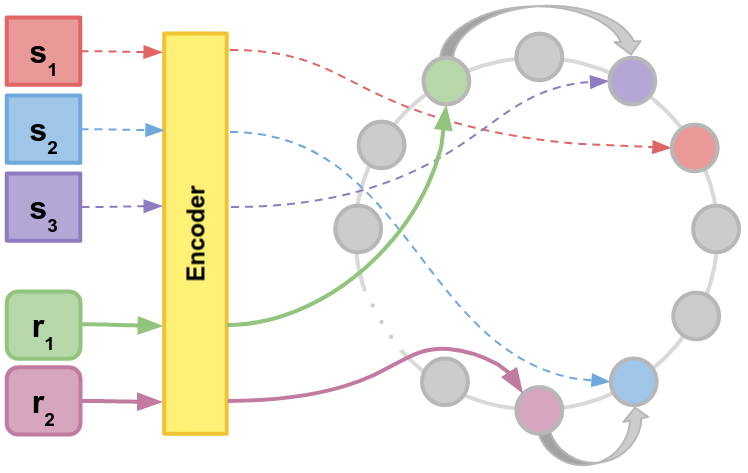}
  \caption{\label{fig:HD_hashing} Illustration of the operation of HD hashing. In this example, after encoding each of the three servers and two requests to a circular-hypervector, $r_1$ is assigned to server $s_3$, which is the server whose hyperspace representation is closest to its. Likewise, $r_2$ is assigned to $s_2$. Note that, unlike consistent hashing, the direction of rotation does not matter.}
\end{figure}

HD hashing, illustrated in Figure~\ref{fig:HD_hashing}, draws inspiration from consistent and rendezvous hashing, but seeks a solution that is both robust and efficient by translating the problem into a hyperdimensional computing task.

Let $S=\{s_1 ,\dots,s_k \}$ be a set of $k$ servers, $R=\{r_1 ,\dots,r_\ell \}$ a set of $\ell$ requests and $C=\{\mathbf{c}_1 ,\dots,\mathbf{c}_n \}$ a set of $n>k$ hypervectors. We also denote by $h(\cdot)$ a hash function that takes as input a server or request. The process of adding servers to the system in HD hashing is similar to consistent hashing, but instead of mapping them to a unit interval (see Sec.~\ref{sec:consistent_hashing}), HD hashing assigns (or "\textit{encodes}" in HDC terminology) each server to a hypervector. To distribute requests among servers, HD hashing also encodes each request. Let us represent this encoding by the function $\mathrm{Enc}: S\cup R \to C$. Then, HD hashing encodes every server and request as follows:
\begin{equation} \label{eq:encoding}
    \mathrm{Enc}\left(x\right)\;=\;C\big[h(x)\bmod n\big]
\end{equation}
where $x$ is either a server or a request and $C\left[h(x)\bmod n\right]$ denotes the hypervector at position $h(x)\bmod n$ in $C$. 

With all servers and requests encoded to the hyperspace, each request $r_i$ is mapped to server $s_j$, such that:
\begin{equation} \label{eq:inference}
    s_j\;=\;\underset{s \in S}{\arg\max} \;\delta\big(\mathrm{Enc}(s), \mathrm{Enc}(r_i )\big)
\end{equation}
where $\delta$ is a given similarity metric between a pair of hypervectors such as inverse Hamming distance or the cosine similarity. The operation above is the one mentioned in Section~\ref{sec:HDC}, and it is called inference due to the first applications of HDC in learning tasks.
% Motivated by the first applications of HDC in learning tasks, the operation above is the one mentioned in Section~\ref{sec:HDC}, called inference.
This is exactly the operation that Schmuck et al.~\cite{schmuck2019hardware} show to be optimizable to the extreme of a single clock-cycle in special hardware. In other words, by using hardware accelerators for HDC each mapping in HD hashing could be executed in $\mathcal{O}(1)$ time.

One remaining, but crucial, question is: how do we create the set of hypervectors $C$? Similar to consistent hashing, we map servers and requests onto a circle. We then map the request to the server that is assigned to the nearest node on the circle according to Eq.~\ref{eq:inference}. To accomplish this, we introduce \textit{circular-hypervectors} as a way of representing a circle in hyperspace such that the closer a node is on the circle the more similar its hypervector. More properties of circular-hypervectors and the process to create them are described in the next section.
\section{Circular-Hypervectors}
\label{sec:circular_encoding}

To understand circular-hypervectors we first describe random and level-hypervectors, both types are used to represent information in hyperspace, a process called \textit{encoding}. Encoding strategies have already been proposed for various types of input data, such as images~\cite{manabat2019performance}, time series~\cite{imani2017voicehd} and text~\cite{najafabadi2016hyperdimensional}. The process usually starts by generating a set of randomly sampled hypervectors that represent discrete atomic pieces of information (e.g. discretized amplitudes of a signal, values of a feature, symbols or identifiers). From these so-called basis-hypervectors more complex objects like the ones listed above can be encoded by combining and manipulating the basis-hypervectors using bundling, binding and permutation operations. 

The basis-hypervectors can be correlated with each other depending on what they represent. For example, consider temperature. Clearly there is a stronger correlation between closer temperatures. On the other hand, for symbols such as letters, this correlation does not necessarily exist. Naturally, the most successful encoding techniques are able to translate these correlations into hyperspace. For this reason, categorical data (letters for example) are encoded with independently and uniformly sampled \textit{random-hypervectors}, while scalar information (e.g. temperature) is represented using \textit{level-hypervectors}~\cite{rahimi2016hyperdimensional}. 

Level-hypervectors are created by quantizing an interval to $m$ levels and assigning a hypervector to each. The similarity between hypervectors is proportional to the distance between the intervals. This correlation is achieved by assigning a random $d$-dimensional hypervector to the first interval, and after this, subsequent intervals are obtained by flipping $d/m$ random bits at each interval. As a result, the last hypervector is completely dissimilar to the first one.

\begin{figure}
 \centering
 \includegraphics[width=\columnwidth]{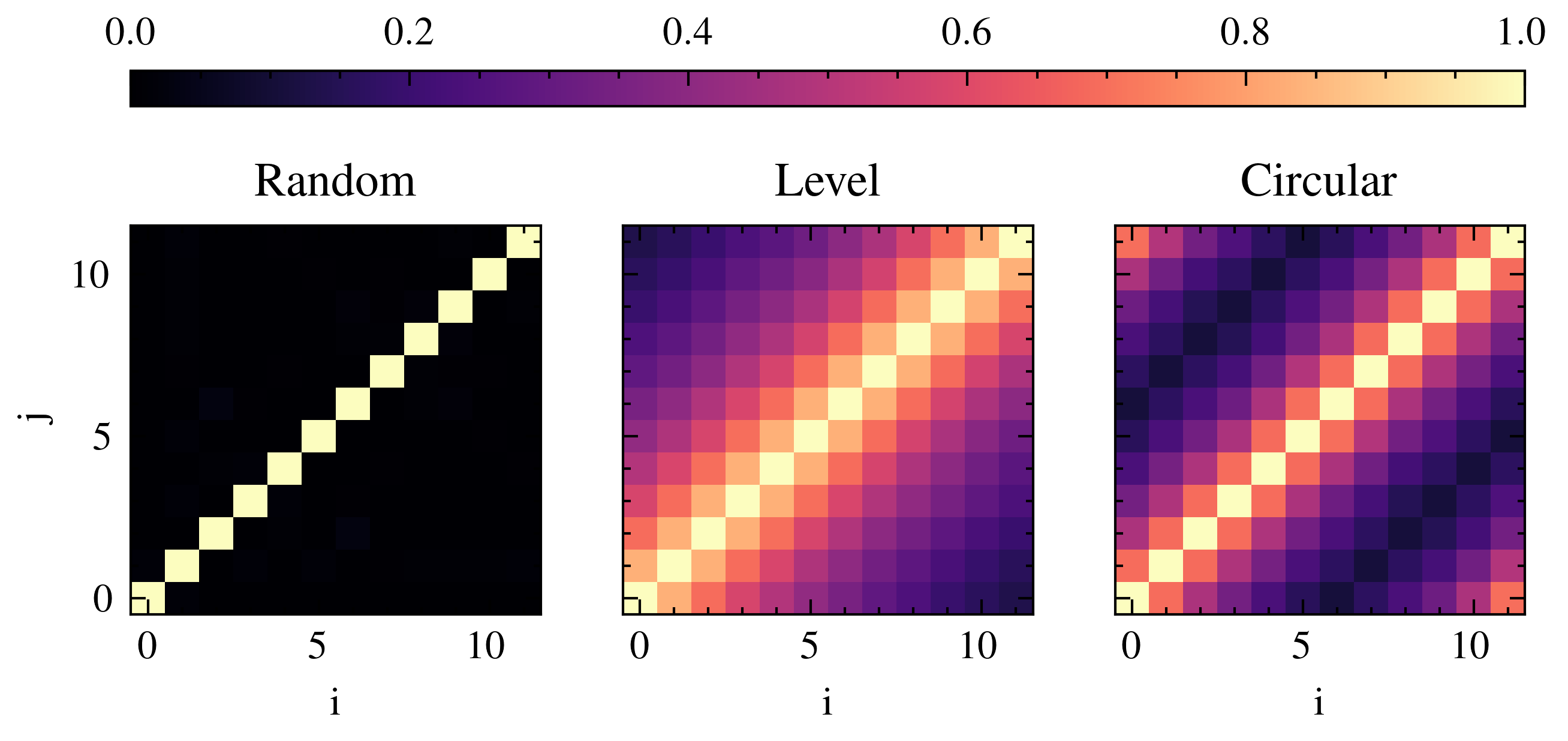}
 \includegraphics[width=\columnwidth]{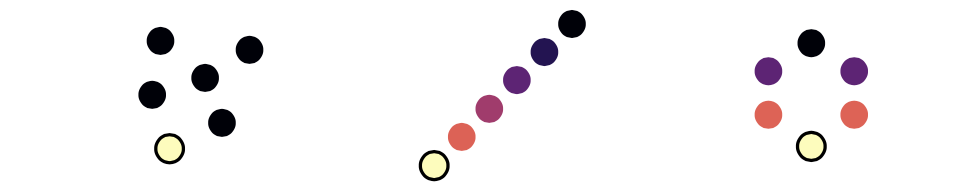}
  \caption{\label{fig:periodic-distances} Pairwise cosine similarities between hypervectors $i$ and $j$ within different sets of 12 basis-hypervectors. An alternative visualization in which each hypervector is represented by a node is shown below. The colors indicate the similarity with the yellow reference node.}
\end{figure}

Circular-hypervectors are an extension to level-hypervectors that eliminate the discontinuity in similarity between the last and first interval, as visualized in the similarity profiles in Figure~\ref{fig:periodic-distances}. By removing the discontinuity, the hypervectors become a set with circular correlation.

\begin{figure}[h]
 \centering
 \includegraphics[width=.75\columnwidth]{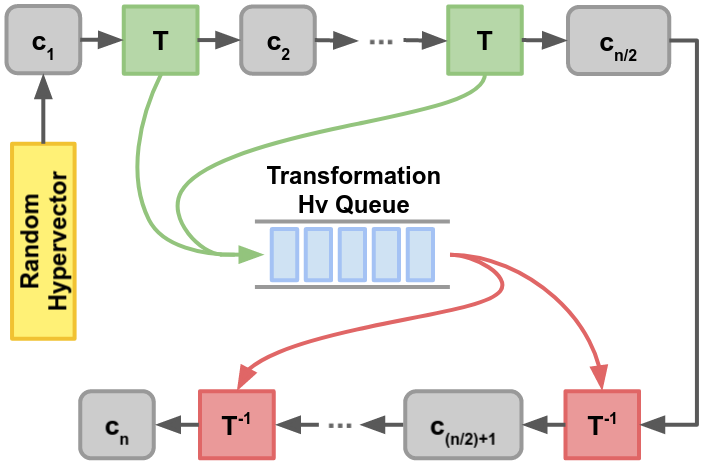}
  \caption{\label{fig:periodic-hvs} Illustration of the process to create circular-hypervectors. The curved arrows represent transformation-hypervectors being inserted in/removed from $Q$.}
\end{figure}

The procedure for generating circular-hypervectors, illustrated in Figure~\ref{fig:periodic-hvs} and detailed in Algorithm~\ref{alg:create_periodic}, starts with a single random-hypervector, uniformly sampled from the hyperspace of dimension $d$ (generated by the function \textit{random\_hypervector($d$)} in the algorithm). From there, inspired by the creation of the level-hypervectors, a sequence of transformations (T) are made to create $n/2$ level correlated hypervectors. Such transformations consist of XORing (also called \textit{binding} in HDC, represented by the symbol $\oplus$) with what we name transformation-hypervectors ($\mathbf{t}$), which are placed in a queue ($Q$). The second half of hypervectors are then obtained by performing backward transformations (T$^{-1}$): the transformation-hypervectors are popped from $Q$ (\textit{first-in-first-out}) and sequentially bound to the current vector in order to generate the next one.

\begin{algorithm}
    \SetKwInOut{Input}{Input}
    \SetKwInOut{Output}{Output}

    \Input{Two integers $n$ and $d$.\footnotemark}
    \Output{A set $\{\mathbf{c}_1 ,\dots,\mathbf{c}_n \}$ of $n$ $d$-dimensional circular-hypervectors.}
    
    Define an empty queue $Q$ \hfill\tcp{Transformation Hv Queue}
    $\mathbf{c}_1 \gets$  random\_hypervector($d$)\\
    \tcc{Perform forward transformations (T)}
    \For{$i\in\{2, \ldots, \frac{n}{2}\}$}
    {
        $\mathbf{t} \gets \mathbf{0}^d$
        \hfill\tcp{$d$-dimensional zeros vector}
        Flip $d/m$ random bits of $\mathbf{t}$\\
        $\mathbf{c}_i \gets \mathbf{c}_{i-1}\oplus \mathbf{t}$\\
        Enqueue$(Q,\mathbf{t})$
    }
    \tcc{Perform backwards transformations (T$^{-1}$)}
    \For{$i \in \{\frac{n}{2}+1,\dots,n\}$}
    {
        $\mathbf{t} \gets$ Dequeue$(Q)$\\
        $\mathbf{c}_i \gets \mathbf{c}_{i-1}\oplus \mathbf{t}$
    }
    {
        \Return $\{\mathbf{c}_1 ,\dots,\mathbf{c}_n \}$
    }
    \caption{Creation of circular-hypervectors}
    \label{alg:create_periodic}
\end{algorithm}
\footnotetext{For ease of understanding, this version assumes that $n$ is even. To generate a set of odd cardinality of circular-hypervectors, simply generate $2n$ and return just $\{\mathbf{c}_1 ,\mathbf{c}_3 ,\mathbf{c}_5 ,\dots,\mathbf{c}_{2n} \}$.}
\begin{figure}[ht]
 \centering
 \includegraphics[width=0.9\columnwidth]{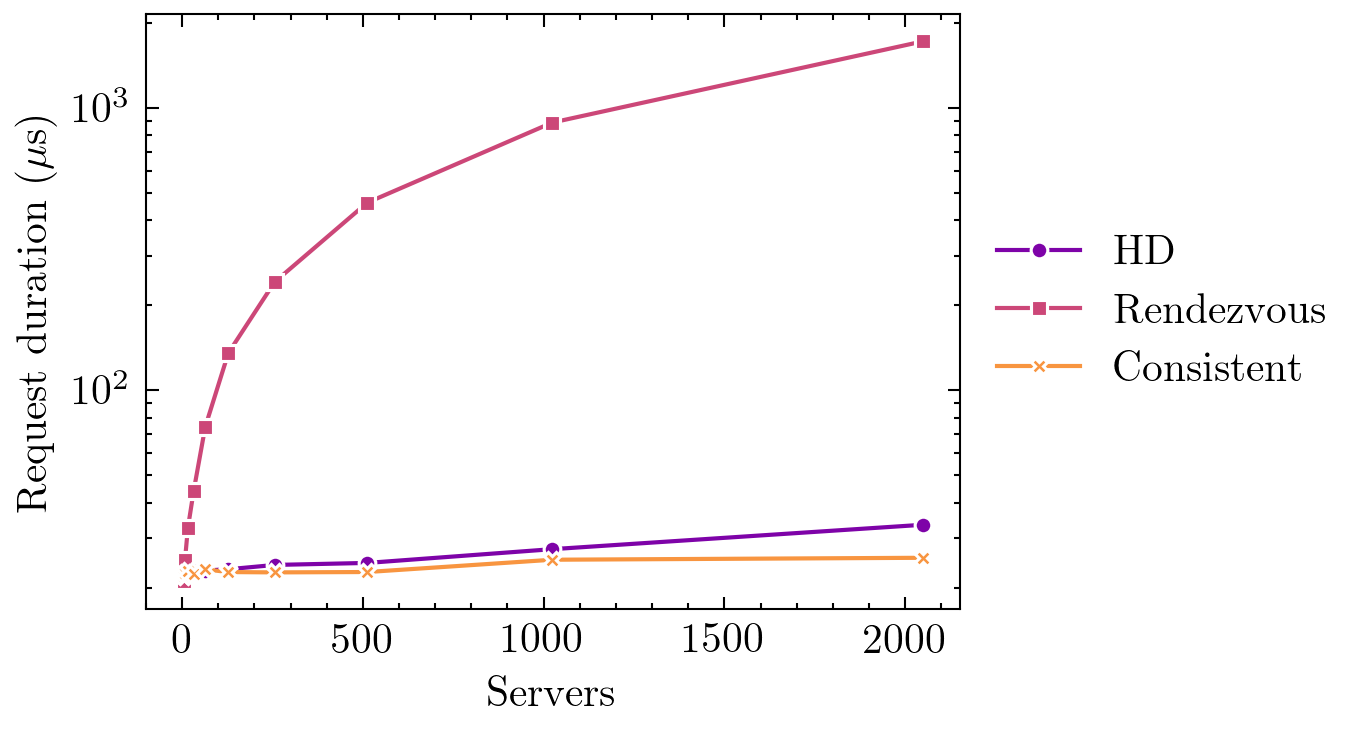}
  \caption{\label{fig:timing} Average request handling duration as the number of servers in the pool increases.}
\end{figure}

\begin{figure*}[ht]
 \centering
 \includegraphics[width=0.95\textwidth]{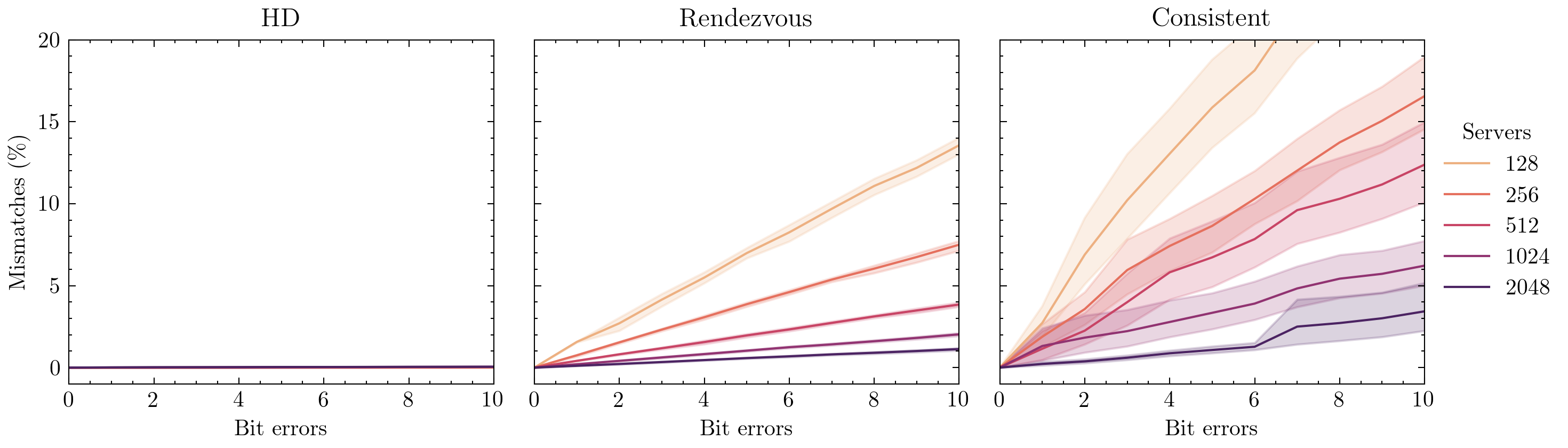}
  \caption{\label{fig:robustness} Percentage of mismatched requests when a number of bit errors occur.}
\end{figure*}

\section{Results and Discussion}
\label{sec:results}

\subsection{Experimental setup}
\label{sec:experimental_setup}
We have created a purpose build emulation framework to empirically verify our results. The emulator consists of two modules, a hash table and a generator. The generator emulates the requests from the outside world being sent to the hash table. The hash table module reads incoming requests from a buffer and uses a hashing algorithm to map them to an available server. Servers are added and removed using two special case requests, a \texttt{join} and \texttt{leave} request, respectively, with a unique identifier of the server. This functional emulator can be used to determine the computational efficiency of various hashing algorithms as well as their robustness to memory errors as we will describe next.

Since we do not have access to specialized HDC hardware and building the hardware is outside the scope of our work we had to implement the HDC operations using commodity hardware. To closely match the parallel nature of HDC hardware, we decided to implement HDC operations on a GPU. We used an Nvidia TITAN Xp GPU with 3840 cores and 12~GB of memory. The GPU's communication overhead was reduced by performing mappings in batches of 256 requests.

Each test was performed with different numbers of servers in the pool, going up to 2048. This scale is enough to show the results and trends of interest, but it is important to emphasize that like the other methods HD hashing can scale to much larger clusters, and even be used hierarchically (standard way to scale such hashing systems~\cite{wang2009hash,shen2006cycloid}) to handle extremely high numbers of servers.

\subsection{Efficiency}
\label{sec:efficiency}
We executed each hashing function in our emulator to empirically determine its computational efficiency. First the generator sends $n$ \texttt{join} requests to add available servers to the hash table module. Then, the generator sends 10,000 requests and tracks the wall-time. From this we determine the average time to handle a request.

\begin{figure}[ht]
 \centering
 \includegraphics[width=0.9\columnwidth]{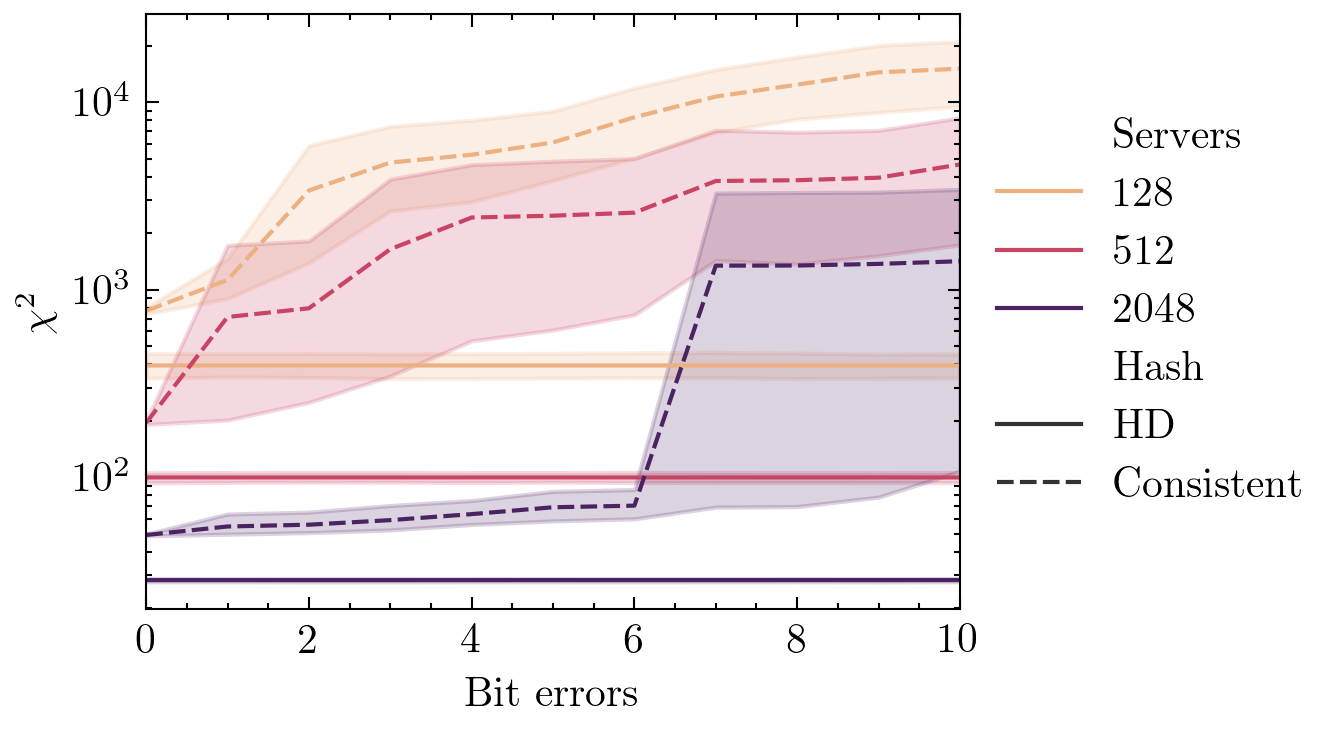}
  \caption{\label{fig:uniformity} The discrepancy between the distribution of requests per server obtained by each algorithm and the uniform distribution, for different numbers of servers and bit errors, measured with the Pearson's $\chi^2$ statistical test.}
\end{figure}

For various numbers of $n$, ranging from 2 to 2048 in powers of 2 the results are shown in Figure~\ref{fig:timing}. The $\mathcal{O}(n)$ time complexity of rendezvous hashing is clearly evident as is the superior computational efficiency of consistent hashing with respect to rendezvous hashing. Our HDC implementation using commodity hardware has a very similar scaling profile to consistent hashing. This confirms our belief that HDC hardware can appropriately be simulated by a GPU. However, as highlighted in Section~\ref{sec:HD_consistent_hashing}, we expect the use of HDC accelerators to reduce the request handling time to a constant with the extreme of a single clock-cycle.

\subsection{Robustness}
\label{sec:robustness}
As motivated before, the other main goal of HD hashing is to be a robust alternative to consistent and rendezvous hashing. In order to assess the performance of each hashing algorithm in an environment subject to noise, two experiments were performed using the emulator's noise injection capabilities. The first and most important, whose results are in Figure~\ref{fig:robustness}, shows how the ability of each technique to map keys to the correct value degrades when a certain number of bits in memory are randomly flipped. Ibe et al.~\cite{ibe2010impact} show that for 22~nm technology, 4-bit and 8-bit bursts occur 10\% and 1\% of the time, respectively. Moreover, errors within a machine are found to be strongly correlated, if a machine experienced an error it is 13-228 times more likely to experience another error in the same month~\cite{schroeder2009dram}. To capture such features of a realistic scenario, we test each hashing technique in the range of 0 to 10 bit flips.

In our experiments, HD hashing confirmed our expectations, turning out to be far superior as none of the requests sent were matched to the wrong server. Meanwhile, in both consistent and rendezvous hashing an increasing percentage of mismatches occur, depending on the noise level.

In the second experiment we tested how uniform the distribution of requests among servers is and how uniform they remain when bits of the hash values are randomly inverted. For evaluation, we used the following \textit{Pearson's chi-squared test}~\cite{greenwood1996guide} to measure goodness of fit between our observed frequency distribution and the uniform distribution:
\begin{align*}
    \chi^2 = \sum_{s_i \in S}\frac{\big(R\left(s_i \right) - E\big)^2}{E}
\end{align*}
where $R(s_i )$ is the number of requests mapped to server $s_i$ by the algorithm and $E = \frac{|R|}{|S|}$ is the uniformity expectation where $|R|$ and $|S|$ are the total number of requests and servers, respectively. The results, illustrated in Figure~\ref{fig:uniformity}, show that not only does HD hashing distribute requests more uniformly than consistent hashing in an ideal scenario, but also that the presence of bit errors worsens the uniformity of consistent hashing even more, while that of HD hashing remains intact. To make the plot more readable, we omit the rendezvous hashing result. Note, from the description of the algorithm in Section~\ref{sec:rendezvous_hashing}, that rendezvous hashing is based only on the output of the hash function, that is, a pseudo-random number. Therefore, its assignment is perfectly (pseudo-) uniform and is not affected by bit errors. Rendezvous hashing, however, still suffers from mismatches and the method has less applicability due to its lower efficiency as illustrated in Figures~\ref{fig:robustness} and~\ref{fig:timing}, respectively.

\section{Future work}
\label{sec:future_work}

Besides being a central component of our work, circular-hypervectors provide a way to represent periodic information that has not been available in the HDC literature thus far. Consider, for example, the seasons of the year, clearly there is a periodic relationship between them.
% Clearly there is a sequential relationship between them, so one might use level-hypervectors to encode them. However, if we start assigning hypervectors from spring the representation of winter would be completely dissimilar. This does not match our intuition since they are chronologically neighboring seasons and share similar characteristics, especially in the transitions between them. 
Several other time-related examples can be listed such as hours of a day or days of a week, as well as other angular data such as directions, geolocation or color spaces. Whether this can be used to improve data representation in HDC, for instance in machine learning applications, is a promising direction of future work.

Our method can utilize the work by Schmuck et al.~\cite{schmuck2019hardware} that shows how HDC accelerators can optimize server lookup (inference in HDC) to a single clock-cycle. Realizing an implementation of the HD hashing algorithm in special hardware is future work.
\section{Conclusion}
\label{sec:conclusion}

We propose Hyperdimensional (HD) hashing---a novel algorithm based on Hyperdimensional Computing (HDC) which allows dynamic scaling of the hash table with minimal rehashing, a problem found in some of the most popular web applications. Through an emulation framework, we compare our method with consistent and rendezvous hashing and the experimental results show that HD hashing is the only approach that guarantees both efficiency and robustness. HD hashing scales similar to consistent hashing, while both are significantly more efficient than rendezvous hashing. Consistent hashing suffers from more than 20\% mismatches with a realistic level of memory errors, which are common in large-scale cloud systems, while HD hashing remains unaffected. This superior level of tolerance to bit errors reduces the chance of critical failures in load balancing and web caching systems, among others.

\section*{Acknowledgment}
This work was supported in part by a UCI Seed grant for Artificial Intelligence Research for Precision Health.

\bibliographystyle{ACM-Reference-Format}
\bibliography{refs_compact}

\end{document}